\documentclass[twocolumn,trackchanges]{aastex631}
\PassOptionsToPackage{draft}{hyperref}  
\usepackage{hyperref}

\usepackage{amsmath}
\usepackage{ulem}

\begin{document}

\title{Intergalactic Wandering Stars in the Local Universe: Theoretical Predictions for Their Distance and Luminosity Distribution}

\author{Jia-Hui Wang}
\affiliation{National Astronomical Observatories, Chinese Academy of Sciences, Beijing 100012, China}
\affiliation{School of Astronomy and Space Science, University of Chinese Academy of Sciences, Beijing 100049, China}
\email{wangjh@bao.ac.cn}

\author{Maosheng Xiang}
\affiliation{National Astronomical Observatories, Chinese Academy of Sciences, Beijing 100012, China}
\affiliation{Institute for Frontiers in Astronomy and Astrophysics, Beijing Normal University, Beijing, 102206, China}
\email{msxiang@nao.cas.cn}

\author{Ji-Feng Liu}
\affiliation{National Astronomical Observatories, Chinese Academy of Sciences, Beijing 100012, China}
\affiliation{School of Astronomy and Space Science, University of Chinese Academy of Sciences, Beijing 100049, China}
\affiliation{Institute for Frontiers in Astronomy and Astrophysics, Beijing Normal University, Beijing, 102206, China}
\email{jfliu@nao.cas.cn}



\begin{abstract}


Intergalactic wandering stars (IWSs) within 10 Mpc remain a poorly explored area of astronomy. Such stars, if they exist, are supposed to be wandering objects as they are not bounded by the gravitational potential of any galaxy. We set out to conduct dedicated studies for unraveling such a wandering stellar population. As the first paper of the series, in the present work we model the distance distribution and luminosity function of IWSs formed via the Hills mechanism of the Galactic central massive black hole (GCMBH). We implement a numerical simulation to generate IWSs taking the ejection history of the GCMBH and the stellar evolution process into consideration, and present their luminosity function in the distance range of 200kpc -- 10Mpc. Our results suggest that a few hundred thousand IWSs have been generated by the GCMBH via the Hills mechanism in the past 14 billion years. These IWSs have an apparent magnitude peaking at 30 to 35\,mag in SDSS $r-$band, which are hard to detect. However, a few thousand of them at the bright end are detectable by upcoming wide-field deep surveys, such as China Space Station Telescope (CSST) and Vera Rubin Observatory (LSST). The forthcoming discovery of such a wandering stellar population will open a door for precise understanding of the matter constitution of the nearby intergalactic space and the dynamical history of galaxies in the local universe. 

\end{abstract}



\section{Introduction} \label{sec:intro}
Intergalactic wandering stars (IWSs) are supposed to be stars unbound to the potential of any host galaxy. Whether or not such stars exist in the local universe, which we define in this paper as within a distance of 10 Mpc, remains an intriguing problem. From the theoretical point of view, IWSs in the local universe should exist, as they can be generated from a number of mechanisms. Apparently, some of them can be born in but later escape from our Milky Way via dynamical effects, for instance, interaction between binary stars and the Galactic central massive black hole (GCMBH) \cite[Hills mechanism;][]{1988Natur.331..687H}.  A detailed mapping of the IWSs in the local universe has at least two implications. First, it is crucial to understand the matter constitution of the nearby intergalactic space. Second, it provides important constraints on the dynamical history of galaxies and groups in the local universe. 


Intergalactic objects have been observed beyond the local universe, particularly inside galaxy clusters.  \citet{1951PASP...63...61Z} first proposed that a diffuse population of intergalactic stars might exist within the Coma cluster through a study of the luminosity function. \citet{1996ApJ...472..145A,1997ApJ...491L..23M,1998ApJ...503..109F,2004ApJ...615..196F} discovered intergalactic planetary nebulae in the Virgo cluster, while \citet{1997MNRAS.284L..11T} discovered intergalactic planetary nebulae in the Fornax cluster. \citet{1998Natur.391..461F} and \citet{2002ApJ...570..119D} detected individual intergalactic RGB stars in the intracluster space within the Virgo cluster. Intergalactic objects within these galaxy clusters are typically believed to originate from tidal stripping within the galaxy cluster \citep{1984ApJ...276...26M} or interactions between galaxies in high-velocity encounters \citep{1976ApJ...204..642R,1996Natur.379..613M}.

There is no reported IWS yet in the local universe. Nonetheless, hypervelocity stars (HVSs) inside the Milky Way have been extensively discovered \citep{2005ApJ...622L..33B, 2005MNRAS.363..223G,2012ApJ...744L..24L,2014ApJ...785L..23Z,2017ApJ...847L...9H,2018ApJ...865...15S,2018A&A...620A..48I,2018ApJ...858....3R,2019ApJS..244....4D,2019A&A...627A.104D,2021A&A...650A.102I}. These stars have enough kinematic energy to escape from the Milky Way and become IWSs only after 
several million years. The escape rate of HVSs due to the Hills mechanism is estimated to be approximately $10^{-4}$ per year \citep{2015ARA&A..53...15B}. Assuming a constant escape rate in the Milky Way's history, we expect that the central massive black hole has produced $O(10^6)$ IWSs over the past ten billion years. Although no confirmed IWSs have been reported, \citet{2012AJ....143..128P} identified intragroup M giant candidates in SDSS data at distances of 300 kpc to 2 Mpc, which may represent such escaping stars or tidally stripped populations. Similarly, \citet{2014ApJ...790L...5B} reported two M giants with estimated distances possibly exceeding 200 kpc, which could be the most distant known Milky Way stars.

The real challenge is how to discover these intergalactic wanderers. First, IWSs are located at large distances, hundreds to thousands of kpc, making them extremely faint for observations. Moreover, the spatial density of IWSs is expected to be low, at a level of about 10 stars per square degree. Identifying these objects therefore needs enormous effort to eliminate contamination from foreground and background objects. Fortunately, upcoming wide-field deep surveys will offer an unprecedented opportunity for IWS detection. In particular, the China Space Station Telescope \citep[CSST;][]{2018cosp...42E3821Z}, the Vera C. Rubin Observatory \citep[LSST;][]{2019ApJ...873..111I}, and the Euclid satellite \citep{2022A&A...662A.112E} will provide deep and multi-band photometry spanning from the ultraviolet to the infrared, which has the potential to identify IWSs.

In light of this, we set out to conduct dedicated studies for unraveling such a wandering stellar population. As the first paper of the series, in this work we present a theoretical estimation of the IWSs distribution in the local universe within 10~Mpc through numerical simulation. Currently, we focus on IWSs produced by the Hills mechanism of the GCMBH. Our numerical simulation combines the Milky Way's potential model \citep{2015ApJS..216...29B} and the stellar evolution model \citep{2016ApJ...823..102C}, allowing us to trace both the star's orbits and its physical properties. Through rigorous probabilistic modeling, we predict the IWSs' distance distribution and luminosity function. 

The paper is organized as follows: Section \ref{sec:model} introduces our numerical models. Section \ref{sec:result} presents the distribution and luminosity function of IWSs, along with the expected detectability for different surveys. Section \ref{sec:conclusion} gives the conclusion.


\section{model}\label{sec:model}

\begin{figure*}
\centering
\includegraphics[width=1\linewidth,]{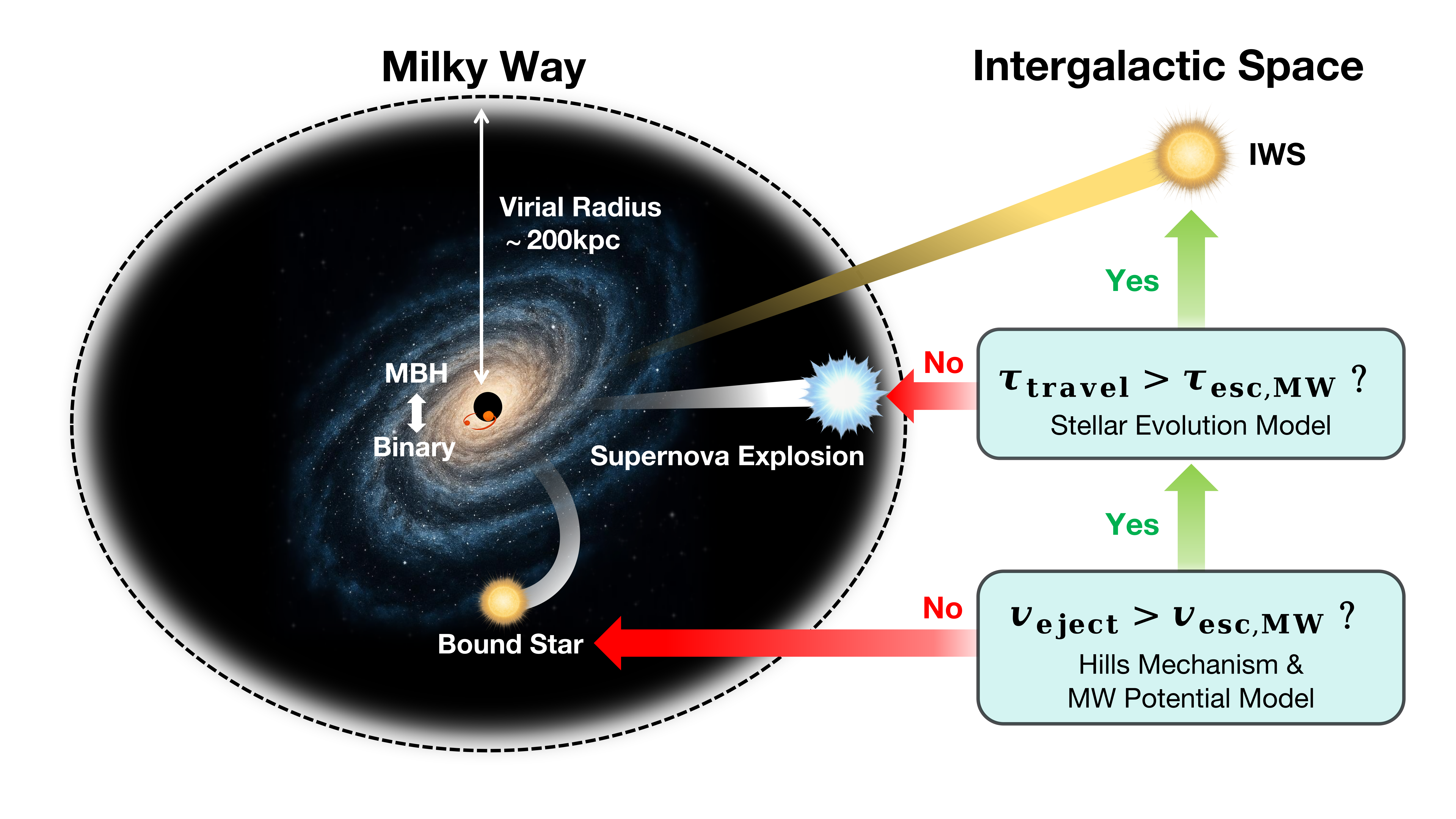}
\caption{Schematic illustration of the IWS formation via the Hills mechanism. When the massive black hole disrupts a binary system, one component may be ejected as a HVS with a velocity $v_{\mathrm{eject}}$ that exceeds the escape velocity of the Milky Way $v_{\mathrm{esc,MW}}$. If the travel time $\tau_{\mathrm{travel}}$ of such an HVS, defined as the cumulative time from its ejection to the present (or until its death), exceeds the time required to escape from the Milky Way $\tau_{\mathrm{esc,MW}}$, the star will ultimately become an IWS.
\label{fig:one}}
\end{figure*}

The formation of intergalactic wandering stars (IWSs) requires two fundamental conditions. First, the stars must reach sufficiently high velocities—enabled by the Hills mechanism—to escape the Milky Way potential, as described by a MW potential model. Second, they must have a sufficiently long stellar lifetime to survive the journey from the Milky Way center to intergalactic space, as constrained by stellar evolution models. These requirements are illustrated in Figure~\ref{fig:one}. Additionally, we propose a probability model to quantify the expected number of observable IWSs. In this section, we introduce them separately.

\subsection{Hills mechanism}\label{sec:hills}

The Hills mechanism provides the ejection velocity of HVSs produced after a binary star is disrupted by a black hole. We adopt the ejection velocity, $v_{\text{eject}}$, as derived from the work of \citet{2006ApJ...653.1194B}:

\begin{equation}
\begin{aligned}
v_{\text{eject}} = &1760 \left( \frac{a_{\text{bin}}}{0.1 \, \text{AU}} \right)^{-\frac{1}{2}} \left( \frac{m_1 + m_2}{2 \, M_\odot} \right)^{\frac{1}{3}} \\  
&\times \left( \frac{M_\bullet}{3.5 \times 10^6 M_\odot} \right)^{\frac{1}{6}} f_R \, \text{km/s}
\end{aligned}
\end{equation}

Here, \( a_{\text{bin}} \) is the binary’s semimajor axis, while \( m_1 \) and \( m_2 \) are the masses of the primary and secondary components of the binary, respectively. \( M_{\bullet} \) is the mass of the GCMBH. \( f_R \) is the adjustment factor for \( v_{\text{eject}} \), and it is expressed as follows:

\begin{align}
    f_R &= 0.774 + \Big( 0.0204 + \Big\{ -6.23 \times 10^{-4} + \Big[ 7.62 \times 10^{-6}  \notag \\
    &\quad + \Big( -4.24 \times 10^{-8} + 8.62 \times 10^{-11} D \Big) D \Big] D \Big\} D \Big) D  \notag
\end{align}

\( D \) is a dimensionless parameter defined by \citet{1988Natur.331..687H}, and it is strongly correlated with the ejection probability, \( P_{\rm eject} \). \citet{2006ApJ...653.1194B} provided a linear approximation of the relationship between \( P_{\rm eject} \) and \( D \):

\begin{equation}
D = \frac{R_{\text{min}}}{a_{\text{bin}}} \left( \frac{2M_\bullet}{10^6 (m_1 + m_2)} \right)^{-1/3}
\end{equation}

\begin{equation}
P_{\text{eject}} \approx 1 - \frac{D}{175} \quad \text{for} \quad 0 \leq D \leq 175
\end{equation}

$R_{\rm min}$ is the closest approach of a binary to the central black hole.

When the binary star masses are unequal, the ejection velocity of the primary component and that of the secondary component are:

\begin{equation}
v_1 = v_{\text{eject}} \left( \frac{2m_2}{m_1 + m_2} \right)^{1/2},
v_2 = v_{\text{eject}} \left( \frac{2m_1}{m_1 + m_2} \right)^{1/2}
\end{equation}

The distribution and range of our model parameters are presented in Table \ref{tab:1}. The black hole mass $M_\bullet$ is set to be \( 4 \times 10^6 \, M_{\odot} \) \citep{2008ApJ...689.1044G}. The primary component's mass distribution follows the Initial Mass Function (IMF), considering both the Kroupa IMF (Model 1) \citep{2002Sci...295...82K} and the Salpeter IMF (Model 2) \citep{1955ApJ...121..161S}, which have different power-law indices for \( m > 1 M_{\odot}\). Previous work has also suggested that the IMF in the Galactic Center may be top-heavy \citep{2010ApJ...708..834B}, which will be discussed in Section~\ref{sec:imf}. The mass range is selected from 0.5 to 15 \( M_{\odot} \), as stars of lower mass are difficult to observe, while stars of higher mass have too short lifetimes to escape the Milky Way and become IWSs. For the secondary component, we assume 20\% equal-mass binaries, where \( m_2 = m_1 \). Observations confirm the prevalence of twin binary stars, with a ratio of 8\% for O-type stars and 30\% for solar-type stars among close binaries \citep{2017ApJS..230...15M}. For the other 80\% of the secondary component, we apply the same IMF sampling as for the primary component \citep{1990MNRAS.242...79T,1991MNRAS.250..701T}. The semi-major axis, $a_{bin}$, has a probability density proportional to \(1/a_{bin}\) \citep{1983ARA&A..21..343A,1980ApJ...242.1063G,1998AJ....115..325H,2006astro.ph..5069K,2007ApJ...670..747K}. We select the range of \( a_{\text{bin}} \) to be 0.05–4 AU, in line with \citet{2006ApJ...653.1194B}, with the lower limit set by the stellar radius and the upper limit by the minimum ejection velocity. The probability density distribution of \( R_{\text{min}} \) is linearly related to \( R_{\text{min}} \) itself, considering gravitational focusing \citep{1988Natur.331..687H}. The range of \( R_{\text{min}} \) is determined by the value of \( P_{\text{eject}} \), which must satisfy \( P_{\text{eject}} > 0 \).

In our study, we set the number of simulations based on the ejection rate of HVSs. The theoretical ejection rate estimated by \citet{1988Natur.331..687H} is $10^{-3}$ to $10^{-4}~\rm yr^{-1}$. \citet{2013ApJ...768..153Z} estimated an ejection rate of $10^{-4}$ to $10^{-5}~\rm yr^{-1}$ by comparing observations of S-stars and B-type HVSs with simulations. \citet{2014ApJ...787...89B} discovered 21 late-type B stars in the HVS survey, inferring an ejection rate of $1.5\times10^{-6}~\rm yr^{-1}$ for stars with masses between 2.5 and 4 \(M_{\odot}\), and a total HVS ejection rate of $10^{-4}~\rm yr^{-1}$. In our work, we calibrate the total number of HVSs based on the observational results of \citet{2014ApJ...787...89B}. Over the age of the Milky Way, which is about 14 billion years, a total of $1.5\times10^{-6}~\rm yr^{-1}\times14\times10^{9}~yr=21,000$ stars with masses between 2.5 and 4~\(M_{\odot}\) would have been ejected. Ejection numbers for stars of other masses are sampled according to the selected IMF. It should be noted that here we assume the ejection rate to be constant over time. The ejection rate is probably a complex function of the star formation rate (SFR), the fraction of close binaries, and the binary orbital parameters. It is possible that the ejection rate was larger in the early history, because some studies have suggested that the SFR may have been higher during that period \citep{1997ApJ...490..493N,2003A&A...399..931Z,2008AJ....136..367M}. In this case, the number of IWSs would exceed our estimates.

\begin{deluxetable*}{lcccc}
\tabletypesize{\scriptsize}
\tablewidth{0pt} 
\tablecaption{Parameter and Model \label{tab:1}}
\tablehead{
\colhead{Parameter} & \colhead{Symbol} & \colhead{Distribution} & \multicolumn{2}{c}{Value or Range} \\
\cline{4-5}
\colhead{} & \colhead{}  & \colhead{} & \colhead{Model 1} & \colhead{Model 2} 
}
\startdata 
{Mass of the GCMBH} & $M_\bullet$ & - & $4\times10^6 M_{\odot}$ & $4\times10^6 M_{\odot}$ \\
{Primary mass} & $m_{1}$ & $P(m_1)\propto m_{1}^{-\alpha}$ & \parbox[t]{3cm}{\centering $\alpha=2.3$; $0.5 < m_1 < 1$\\ $\alpha=2.7$; $1 < m_1 < 15$} & $\alpha=2.35$; $0.5 < m_1 < 15$ \\
{Secondary mass} & $m_{2}$ & - & \parbox[t]{4cm}{\centering $m_{2} = m_{1}(20\%)$ \\ $P(m_2) = P(m_1)$ (80\%)} & \parbox[t]{4cm}{\centering $m_{2} = m_{1}(20\%)$ \\ $P(m_2) = P(m_1)$ (80\%)} \\
{Semimajor axis} & $a_{bin}$ &$P(a_{bin})\propto 1/a_{bin}$& $0.05-4$ AU & $0.05-4$ AU  \\
{Closest distance} & $R_{min}$ &$P(R_{min})\propto R_{min}$ & $1-1500$ AU & $1-1500$ AU \\
\enddata

\tablecomments{Model 1 adopts the Kroupa IMF, while Model 2 adopts the Salpeter IMF.}
\end{deluxetable*}

\subsection{MW potential model}

We calculate the orbits of escaping stars using the \texttt{Gala} Python package \citep{gala,adrian_price_whelan_2020_4159870}, taking the \texttt{MilkyWayPotential} of \citet{2015ApJS..216...29B}. The Milky Way's mass distribution behind this potential model includes a spherical nucleus, a Miyamoto-Nagai disk\citep{1975PASJ...27..533M}, and a spherical Navarro–Frenk–White (NFW) dark matter halo\citep{1997ApJ...490..493N}. 

We examined the escape velocity under this gravitational potential model, requiring stars ejected near the black hole (\(\sim\)1pc) to reach the virial radius of 200~kpc \citep{2006MNRAS.369.1688D}. The results indicate that the ejection velocity must exceed 820~km/s. We conservatively selected 800~km/s as the minimum velocity considered in our simulation.

The precision of orbit calculations is critical. The steep gravitational potential near the central black hole requires a shorter integration time step (\( dt \)) to capture rapid velocity changes, which in turn leads to longer total integration times. We examined different values of \( dt \) and ultimately selected 0.001~Myr, which provides sufficient orbit precision with an affordable time cost.

\subsection{Stellar evolution model}

We adopt the MESA Isochrones \& Stellar Tracks \citep[MIST;][]{2016ApJ...823..102C} to model the evolution of the escaping stars and provide their physical properties and lifetime, $\tau$. To simplify the problem, we fix the metallicity at [Fe/H] = 0. We adopt the MIST grids computed under the initial stellar rotation speed of \( v/v_{\text{crit}} = 0.4 \) \citep[see details in][]{2016ApJ...823..102C}. We perform uniform sampling within the mass range of 0.5–15 \( M_{\odot} \), obtaining a total of 1450 stellar evolutionary tracks through interpolation of the MIST grids. These tracks are resampled at equal time intervals ($\Delta \tau=dt=0.001\;\text{Myr}$), which is used for the probability model calculation in Section 2.4. Simultaneously, we interpolate the magnitudes in the SDSS $u$, $g$, $r$, $i$, and $z$ bands according to the time sampling.

\subsection{Probability model}\label{sec:PM}

To study the spatial distribution of IWSs, we simulate the travel trajectories of millions of escaping stars, each with an escape velocity vector $\vec{v}_{\rm esc}$, whose direction in Galactic coordinate \{$l$,$b$\} is randomly distributed. To reduce the computation cost and speed up the simulation, we propose a probability scheme to estimate the distribution function of the IWSs. 

For an HVS with a given escape velocity vector $\vec{v}_{\rm esc}$, we first calculate all its possible positions $\{\vec{r}\}$ by integrating its orbit assuming it is ejected immediately after its birth, where the position vector refers to the three-dimensional coordinates. In this case, the star travels the full lifetime $\tau$ as an escaping star, and thus reaches the farthest distance consistent with its ejection velocity. In other cases when the star is ejected later by the GCMBH, its travel trajectory as an HVS is a subset of $\{\vec{r}\}$, and we directly retrieved this trajectory subset from $\{\vec{r}\}$ without a recomputation of the orbits. 

For convenience, we discretize $\{\vec{r}\}$ into $N := \tau / \mathrm{d}t$ points. The probability of observing the star at $\vec{r}_{i}$, for $i=1,2,...,N$, is denoted as $P_{\rm obs}(\vec{r}_{i})$:

\begin{equation}\label{eq5}
P_{\rm obs}(\vec{r}_{i}) = \sum_{1}^{N_{{\rm events},i}} P(\tau_{\rm eject}) \times P_{\text{travel}} \times P_{\rm event}(\vec{r}_{i}) 
\end{equation}
the components in the equation are defined as follows:

\begin{itemize}

\item $P(\tau_{\rm eject})$

A star can be ejected at any time during its lifetime, $\tau_{\rm eject}$, depends on when the binary system encounters disruption by the GCMBH. In our simulation, $\tau_{\rm eject}$ is discretized and uniformly sampled with a time interval of 0.001 Myr (=$dt$). This results in $N$ samples of $\tau_{\rm eject}$, thus $N$ possible travel trajectories, as each case of $\tau_{\rm eject}$ sampling results in a unique travel trajectory. We refer to the sampling that results in each of these travel trajectories as an $event$. 

In this work, we assume that each $event$ happens with equal probability, i.e., the star can be ejected at any time throughout its lifetime under the same possibility. The probability of ejection at $\tau_{\rm eject}$, $P(\tau_{\mathrm{eject}})$, is thus given by:
\begin{equation}
P(\tau_{\rm eject})=\frac{1}{N}   
\end{equation}

\item $P_{\text{travel}}$

Our Milky Way has an age of about 14~Gyr. An escaping star with travel time shorter than this may or may not be observed by us now, depending on when the star was ejected in the Milky Way's history $\tau_{\rm eject}$ and its travel time $\tau_{\rm travel}$, which is defined as: 

\begin{equation}
\tau_{\mathrm{travel}} :=
\begin{cases}
\tau - \tau_{\mathrm{eject}} & \text{if } \tau < \tau_{\mathrm{MW}} \\
\tau_{\mathrm{MW}} - \tau_{\mathrm{eject}} & \text{if } \tau \geq \tau_{\mathrm{MW}}
\end{cases}
\end{equation}

Based on the assumption of a constant ejection rate, the probability of observing the star in travel now, \( P_{\text{travel}} \), can be expressed as:

\begin{align}
P_{\text{travel}} &=\frac{\tau_{\rm travel}}{\tau_{\rm MW}- \tau_{\rm eject}} 
\end{align}


\item $P_{\rm event}(\vec{r}_{i})$

For a given $event$, the star can appear at different spatial positions in the trajectory. The total number of possible positions, $N_{\rm position, event}$, is determined by the travel time, $\tau_{\rm travel}$:
\begin{equation}
N_{\rm position, event} = \frac{\tau_{\rm travel}}{dt}.
\end{equation} 

The probability that the star appears at position  $\vec{r}_{i}$ in this $event$, $P_{\rm event}(\vec{r}_{i})$, is thus:
\begin{equation}
P_{\rm event}(\vec{r}_{i}) = \frac{1}{N_{\rm position,event}}
\end{equation}

\end{itemize}

For an HVS with a given escape velocity vector $\vec{v}_{\rm esc}$, the probability of it being observed at position $\vec{r}_{i}$, $P_{\rm obs}(\vec{r}_{i})$, is the summation of all possible $events$ that reach the position $\vec{r}_{i}$, as expressed in Eq.~\ref{eq5}. The number of $events$ that reach $\vec{r}_{i}$, $N_{{\rm events}, i}$, is determined as \( N_{{\rm events},i} = N - i + 1 \).

Taking all these into consideration, and using Eq.~\ref{eq5}, we derive the following equation:

\begin{align}\label{11}
P_{\rm obs}(\vec{r}_{i}) &= \sum_{1}^{N_{{\rm events}, i}} P(\tau_{\rm eject}) \times P_{\text{travel}} \times P_{\rm event}(\vec{r}_{i}) \notag \\
&=\sum_{1}^{N_{{\rm events}, i}}\frac{1}{N} \times 
\frac{\tau_{\rm travel}}{\tau_{\rm MW}- \tau_{\rm eject}} \times 
\frac{dt}{\tau_{\rm travel}} \notag \\
&=\sum_{1}^{N_{{\rm events}, i}}\frac{1}{N} \times
\frac{dt}{\tau_{\rm MW}- \tau_{\rm eject}} 
\end{align}

Numerical calculations demonstrate that the probability of the star being observed in the entire $\{ \vec{r}_i \}$ space is smaller than or equal to 1, as expected.

At this point, for a single escaping star, we can compute the probability of it being observed at each discrete spatial position \(\vec{r}_i\). As mentioned above, we simulate the travel trajectories for a large number of escaping stars to count the number of IWSs. The number of simulations is determined by the number of observed HVSs between 2.5 and 4~$M_\odot$, as introduced in Sect.~2.1.

\section{Result} \label{sec:result}

In this section, we present the distance and luminosity distribution of IWSs in the simulation. In order to understand how the simulation results depend on the model assumptions of initial mass function (IMF), we have implemented two sets of simulations by adopting the Kroupa IMF \citep{2002Sci...295...82K} (Model 1) and Salpeter IMF \citep{1955ApJ...121..161S} (Model 2), respectively (see Table \ref{tab:1} for details). The discussion below will be based on the simulation results of Model 1, but results of the two models will be compared at the end of the section.

\subsection{Distance and luminosity distribution of IWSs} \label{sec:result1}

\begin{figure}
\centering
\includegraphics[width=0.95\linewidth,]{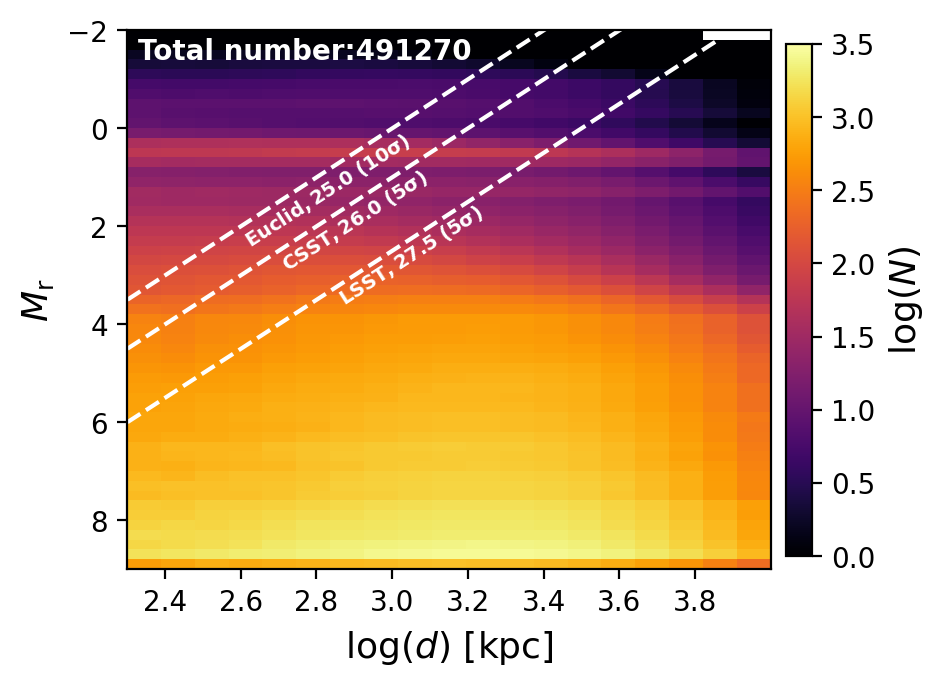}
\caption{The distribution of IWSs in the luminosity-distance space. The figure shows the distribution of IWSs within the range of 200kpc–10Mpc. Color represents the logarithmic number density of IWSs, and the total number is 491,270. The dashed lines indicate the observation limits for CSST, LSST, and Euclid surveys. 
\label{fig:2}}
\end{figure}

The Figure~\ref{fig:2} shows the number density distribution of IWSs in the luminosity–distance space. Here we adopt absolute magnitude in the SDSS $r$-band as an indicator of luminosity, and define an HVS star with Galactocentric distance larger than 200~kpc as an IWS. Our statistics are focused on IWSs with distances from 200~kpc to 10~Mpc, and ignore further objects. The density increases with increasing $M_r$, which is a natural consequence of the IMF that fainter, lower mass stars dominate over brighter, more massive stars. Stars more massive than 5~$M_\odot$ cannot contribute to the IWS population as their lifetime is too short to allow them to escape from the Milky Way. 

In the distance dimension, the IWSs' number density seems to exhibit a concentration in the regime of $3.2 < \log~d < 3.6$~kpc. This is a joint consequence of two effects. At $\log~d \lesssim 3.4$~kpc, the increase in density with increasing $\log~d$ is due to the increase in the volume corresponding to the logarithmic distance bin, while the actual number of stars decreases with increasing distance in linear space. At $\log~d \gtrsim 3.4$~kpc, the decrease in density with increasing $\log~d$ is, however, mainly due to a dramatic decrease in the underlying stellar density, as only a small portion of IWSs have a long enough lifetime and have experienced sufficient travel time to reach such a large distance.

The figure also marks the detection limits of a few upcoming large surveys based on their designed limiting magnitudes, namely CSST \citep[r=26.0, 5$\sigma$;][]{TB-2021-0016}, Euclid \citep[r=25.0, 10$\sigma$;][]{2025A&A...697A...2E}, and LSST \citep[r=27.5, 5$\sigma$;][]{2022ApJS..258....1B}. It illustrates that while most of the IWSs are too faint to be observed by these surveys, a portion of IWSs with modest distance at the bright end may have a chance to be detected, especially those with spectra and high signal-to-noise ratio. It can be seen that the CSST limit encloses a substantial number of main-sequence (turnoff) IWSs with $2 < M_{\rm r} < 4$ in the distance regime within $\sim$1~Mpc. Notably, the number density exhibit a distinct peak at $M_{\rm r} \simeq 0.5$, which is attributed to the numerous red clump stars in the H-R diagram.

\begin{figure}
\centering
\includegraphics[width=0.95\linewidth,]{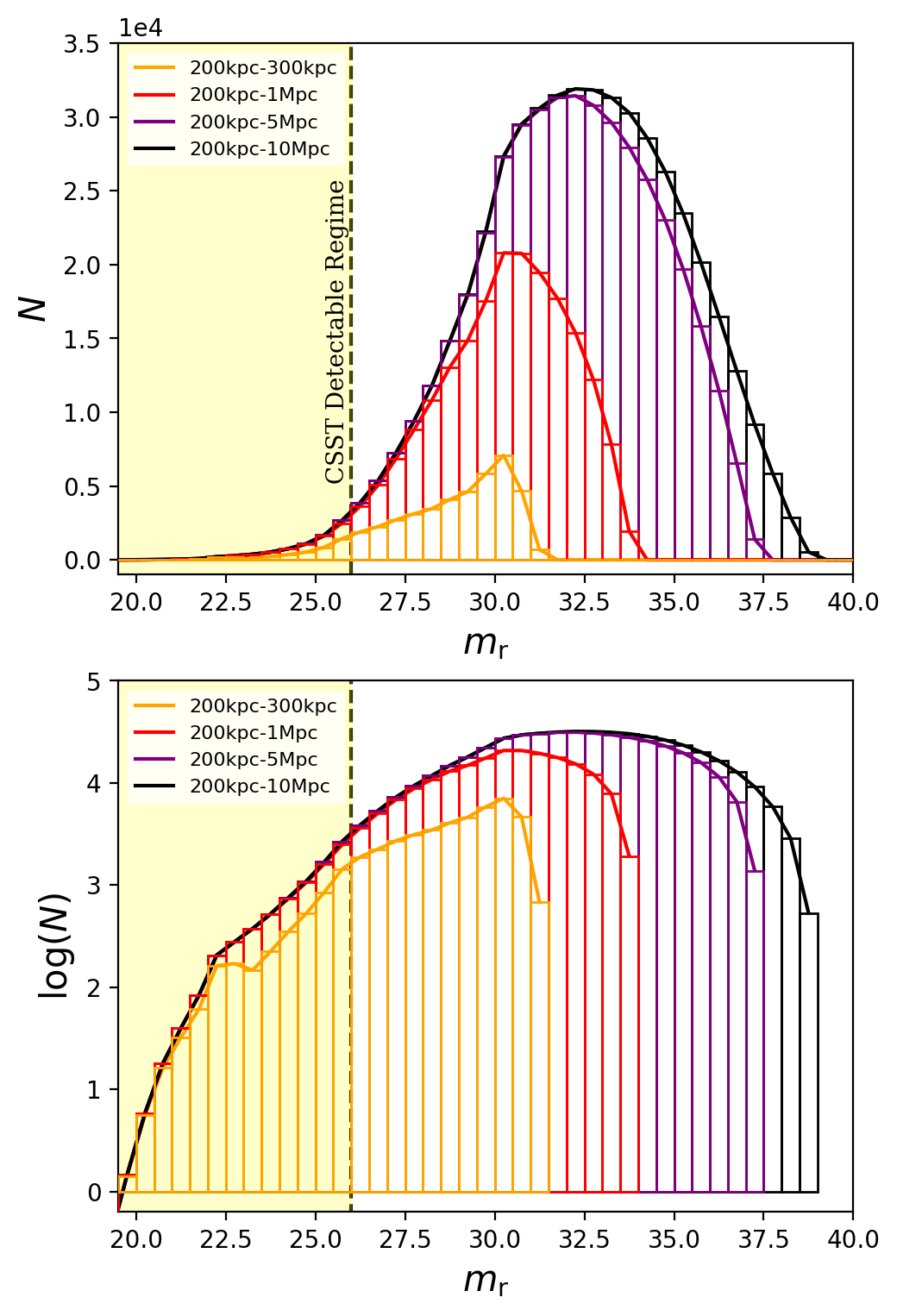}
\caption{Luminosity function of IWSs produced via the Hills mechanism of the GCMBH. The upper panel shows the results for IWSs in four different distance regimes, as marked in the figure. The yellow shaded region indicates the detectable regime of the CSST survey. The lower panel shows the same distributions but with the y-axis in logarithmic scale.
\label{fig:3}}
\end{figure}

\begin{figure*}[ht!]
\centering
\includegraphics[width=0.9\linewidth,]{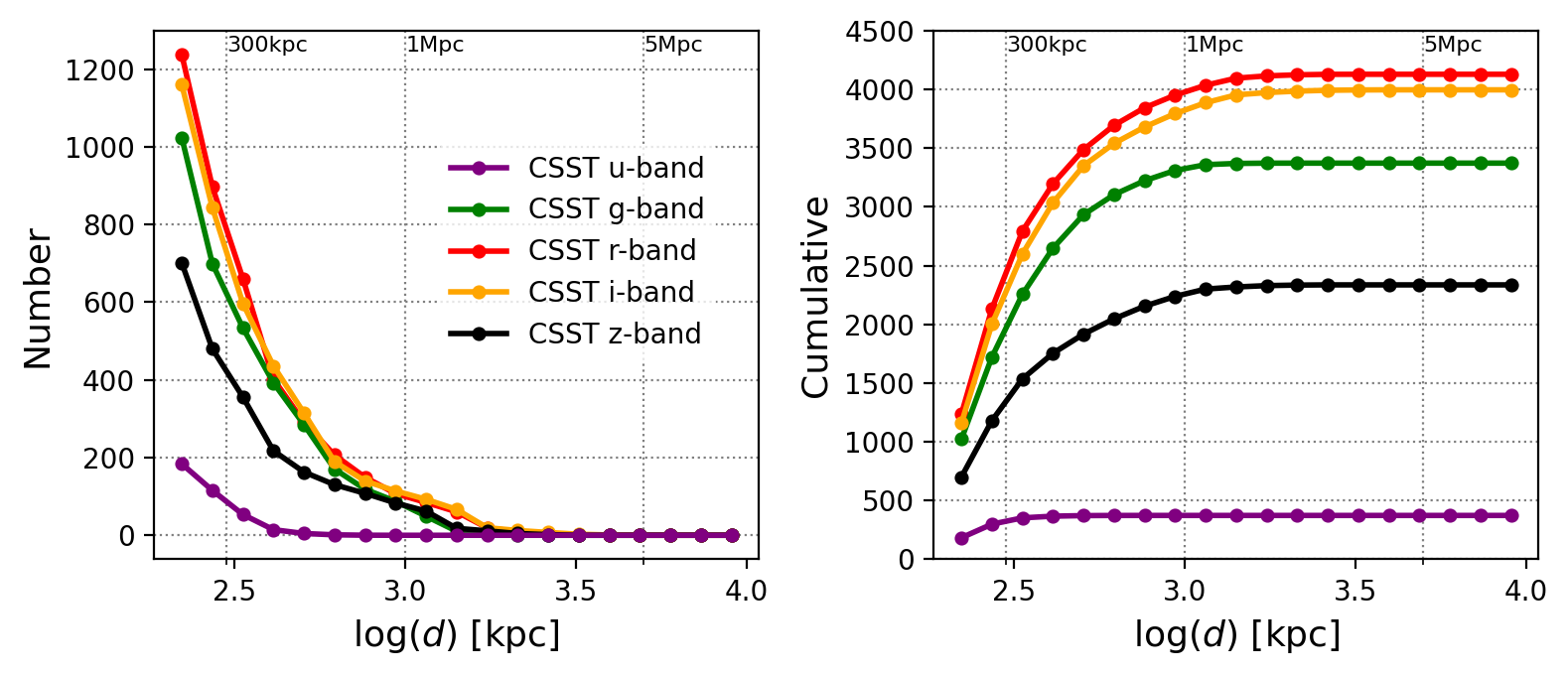}
\caption{Distance distribution of expected IWSs in the CSST survey. The left panel shows the number of IWSs in 19 logarithmic distance bins, uniformly spaced from $\log(d)$ = 2.3 ($d$ = 200kpc) to $\log(d)$ = 4.0 ($d$ = 10Mpc), with different colors representing detection for different filter bands. The right panel shows the cumulative number of IWSs as a function of distance.
\label{fig:4}}
\end{figure*}

\begin{figure*}[ht!]
\centering
\includegraphics[width=0.9\linewidth,]{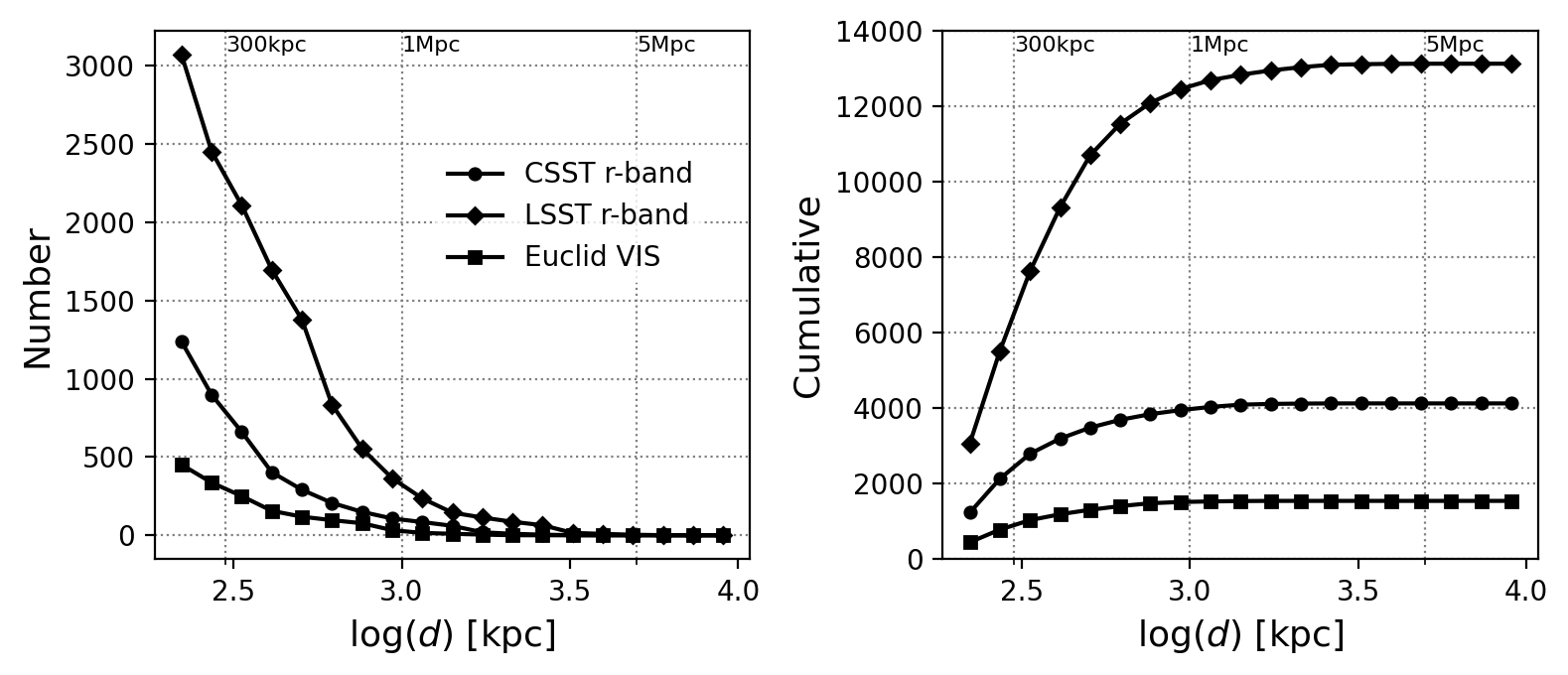}
\caption{Same as Figure \ref{fig:4}, but for comparison among CSST, LSST, and Euclid.
\label{fig:5}}
\end{figure*}

Figure~\ref{fig:3} shows the luminosity function of IWSs in different distance intervals. In the 200–300~kpc distance interval, the SDSS $r$-band apparent magnitude exhibits a gradual rise until peaking at $m_{\rm r} \simeq 30$. As the distance interval increases, the luminosity peak shifts toward fainter magnitudes, reaching $m_{\rm r} \simeq 33$ in the 200kpc–10Mpc interval. This trend is a reflection of two effects: first, stars of the same intrinsic luminosity appear fainter at greater distances; second, only faint, low-mass stars possess sufficiently long $\tau_{\rm travel}$ to reach further intergalactic space. As a result, as distance increases low-mass IWSs become increasingly dominant. The lower panel also clearly shows that a minor but visible population of IWSs at the bright end can be brighter than 26~mag within the observation limit of CSST ($m_{\rm r}<26$).

\subsection{Prediction for CSST and other large surveys}

CSST is a 2-meter space survey telescope with a field of view of $1.1^\circ \times 1.2^\circ$. It will survey a large sky area of 17,500 square degrees, obtaining multi-band photometry and slitless spectra in the full wavelength range of 255–1000~nm for objects down to 26~mag ($5\sigma$) in $r$-band \citep{TB-2021-0016} (~23.4 mag for spectra). These deep, multi-band photometry and slitless spectra with ultraviolet coverage will provide unprecedented opportunities for identifying IWSs.


Figure~\ref{fig:4} shows our model prediction for the number of IWSs in the CSST photometric survey as a function of distance. The results are shown for five filter bands: $u$, $g$, $r$, $i$, and $z$. A total of 3500 to 4000 IWSs are detectable in CSST $g$-, $r$-, and $i$-bands, and most of them have a distance within 2~Mpc. However, only about 400 IWSs are detectable in the $u$-band, mostly within 500~kpc. This is mainly because most of the IWSs are intrinsically fainter in the ultraviolet.

\begin{figure}
\centering
\includegraphics[width=0.95\linewidth]{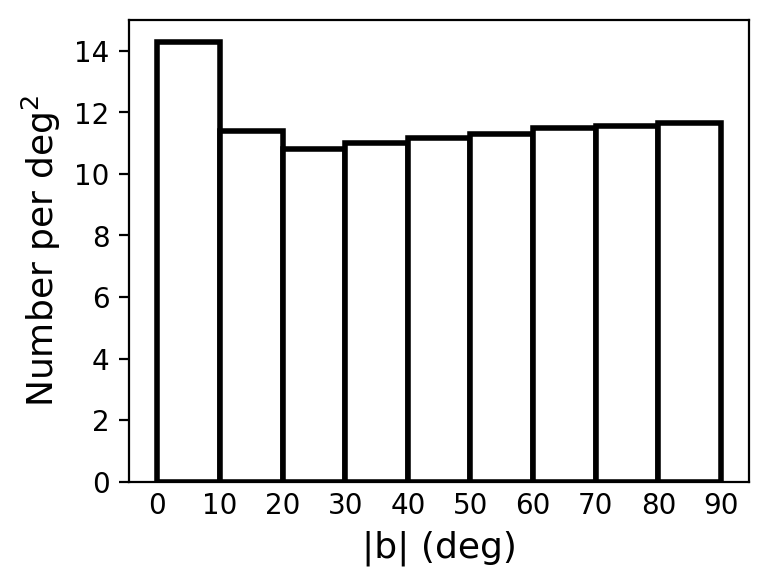}
\caption{Number density distribution of IWSs in our simulation as a function of Galactic latitude.
\label{fig:66}}
\end{figure}


Figure~\ref{fig:5} compares the model predictions of IWSs' number density distribution among the CSST, LSST, and Euclid surveys. We only show results for the $r$-band as it matches the Euclid VIS  filter \citep[550--900~nm;][]{2024arXiv240513492E} (similar results are obtained for the $i$-band). It shows that the LSST survey may detect the largest number of IWSs ($\sim13,000$), as it has the deepest limiting magnitude. Most of these IWSs have a distance within 3~Mpc. The Euclid survey may detect about 1,500 IWSs, most of which are within 1~Mpc. 

Our simulations reveal a nearly uniform distribution of IWSs across Galactic latitude, with only a slight excess at low latitudes (\(|b| < 10^\circ\)) due to the influence of the Galactic disk, as shown in Figure 6. In light of this, when counting the number of IWSs that are detectable by these surveys, we have simply based our estimate on the total number of IWSs in the simulation and the fractional sky area, instead of drawing the actual footprint of the surveys. We note that our simulation results are achieved based on the assumption that the $\vec{v}_{\rm esc}$ is randomly distributed in spatial directions, while the true spatial distribution of HVSs ejected by the central massive black hole remains poorly understood \citep{2015ARA&A..53...15B}. Any preference for the direction of $\vec{v}_{\rm esc}$ may introduce different results to the current simulation. Nonetheless, as these surveys will observe a large portion (about a half) of the sky, we expect that our results stay robust in order of magnitude.

It is important to note that numbers presented in Figure~4 and Figure~5 only refer to IWSs that can be detected but not necessarily identified by the surveys. To identify IWSs from the surveys is a much harder task, as it requires a precise classification of their stellar nature, as well as a reliable determination of their physical parameters, especially their distance. At this point, the CSST, LSST, and Euclid surveys will provide multi-band photometric data covering wavelengths from the ultraviolet to the infrared, which will be helpful for classification. Meanwhile, the ultraviolet photometry and ultraviolet slitless spectra from CSST may provide robust stellar parameters for several hundred of IWSs at the brightest end.

\begin{figure}
\centering
\includegraphics[width=0.95\linewidth,]{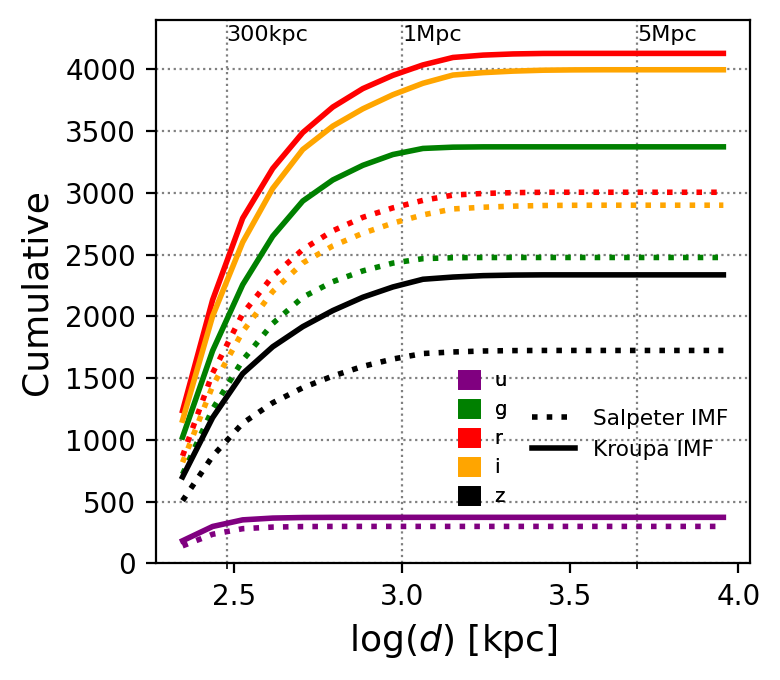}
\caption{Cumulative distance distribution of IWSs predicted for the CSST survey. Solid lines are simulation results based on the Kroupa IMF, while dashed lines are results based on the Salpeter IMF. Different colors are predictions for the CSST \(u\), \(g\), \(r\), \(i\), and \(z\) bands. 
\label{fig:6}}
\end{figure}

\subsection{Impact of the initial mass function} \label{sec:imf}

While we have shown results for models adopting the Kroupa IMF, models with other forms of IMF may lead to moderately different IWS distributions. Figure~\ref{fig:6} shows a comparison of the IWS distribution between the model adopting the Kroupa IMF (Model~1) and the model adopting the Salpeter IMF (Model~2). Model~2 predicts a smaller number of IWSs, about 70\% of Model~1, a natural consequence of the fact that the Kroupa IMF has a larger fraction of lower mass stars.

However, the actual Galactic central IMF might be different from both of these, for example due to the influence of the Galactic central environment or mass transfer in binaries. It has been suggested that the Milky Way center may have an extremely top-heavy IMF ($\alpha\simeq0.45$) at the high-mass end ($M\gtrsim7$~$M_\odot$) \citep[e.g.][]{2010ApJ...708..834B}. However, as discussed in Section~\ref{sec:result1}, massive HVSs are unlikely to become IWSs as their lifetime is too short to reach the outskirts of the Milky Way. We therefore expect that our results for IWSs are not seriously affected by an extremely top-heavy IMF at the high-mass end.

\section{Conclusion} \label{sec:conclusion}


Intergalactic wandering stars (IWSs) remain an unexplored population of the nearby universe. In this work, we model IWS that escaped from the Milky Way via the Hills mechanism of the Galactic central massive black hole. Our numerical simulation shows that a few hundred thousand HVSs have escaped from our Milky Way and become IWSs in the past 14 billion years. Most of these IWSs have reached a distance between 200~kpc and a few Mpc. Their apparent magnitudes exhibit a peak between 30 and 35\,mag in SDSS $r$-band, depending on the distance. Several to tens of thousands of these IWSs can be brighter than $\sim$26~mag, within the limiting magnitude of the upcoming large surveys. In particular, we expect that the CSST survey will achieve a breakthrough in identifying IWSs, owing to its powerful ability to deliver robust stellar parameters from the ultraviolet photometry and spectroscopy. This will open a door for a better understanding of the matter constitution of the nearby intergalactic space as well as the dynamical history of galaxies in the local universe.

We note that the Hills mechanism discussed in this study is only one of the possible formation mechanisms of IWSs. Other mechanisms, such as supernova explosions and tidal interactions with dwarf galaxies, are also expected to produce IWSs. It is likely that stellar population in the intergalactic space of the local universe is more complex, and we plan to conduct extensive follow-up studies in future.

\noindent {\bf Acknowledgments}
We acknowledge financial support from the National Key R\&D Program of China grant No. 2022YFF0504200 and NSFC grant Nos. 2022000083. 

\bibliography{sample631}{}
\bibliographystyle{aasjournal}



\end{document}